\documentclass{article}
\usepackage{spconf,amsmath,bm,diagbox,subcaption}
\usepackage{graphicx}
\usepackage{hyperref}
\usepackage{color}
\usepackage{soul}
\hypersetup{
colorlinks = true,
linkcolor = blue,
citecolor = blue
}

\DeclareMathOperator*{\argmax}{arg\,max}

\title{Unpaired Image Denoising}

\name{Priyatham Kattakinda$^{*}$, A. N. Rajagopalan\thanks{* Corresponding author, email id: priyathamkat@smail.iitm.ac.in}}
\address{Dept. of Electrical Engineering, IIT Madras}

\begin{document}

\onecolumn
\noindent © 2020 IEEE.  Personal use of this material is permitted.  Permission from IEEE must be obtained for all other uses, in any current or future media, including reprinting/republishing this material for advertising or promotional purposes, creating new collective works, for resale or redistribution to servers or lists, or reuse of any copyrighted component of this work in other works.
\twocolumn
	
\pagebreak

\maketitle

\begin{abstract}
Deep learning approaches in image processing predominantly resort to supervised learning. A majority of methods for image denoising are no exception to this rule and hence demand pairs of noisy and corresponding clean images. Only recently has there been the emergence of methods such as \textit{Noise2Void}, where a deep neural network learns to denoise solely from noisy images. However, when clean images that do not directly correspond to any of the noisy images are actually available, there is room for improvement as these clean images contain useful information that fully unsupervised methods do not exploit. In this paper, we propose a method for image denoising in this setting. First, we use a flow-based generative model to learn a prior from clean images. We then use it to train a denoising network without the need for any clean targets. We demonstrate the efficacy of our method through extensive experiments and comparisons.
\end{abstract}
\begin{keywords}
image denoising, flow-based models, unsupervised methods
\end{keywords}
\section{Introduction}
\label{sec:introduction}

Noise corrupts virtually any image captured through a camera. The degradation due to noise is typically captured in the equation: $\pmb{Y} = \pmb{X} + \pmb{N}$ where $\pmb{X}$ is a clean image, $\pmb{N}$ is noise and $\pmb{Y}$ is the corresponding noisy version of $\pmb{X}$. Image denoising methods attempt to recover the clean image from its noisy version.

\noindent Traditional methods such as BM3D \cite{4271520}, NSCR \cite{dong2012nonlocally}, WNNM \cite{gu2014weighted} rely on the self-similarity of image patches to denoise solely from noisy images.
\noindent Methods such as \cite{zhang2017learning, 7839189, lefkimmiatis2017non} that use deep learning have been proposed for image denoising. Although they achieve state-of-the-art performance along with excellent test times, they are all discriminative models. As a result, they require pairs of noisy images and their corresponding clean images.

\noindent Recently, deep learning methods like Noise2Noise \cite{pmlr-v80-lehtinen18a} and Noise2Void \cite{8954066} have been proposed that use statistical properties of noisy image patches to eliminate noise. While these methods do not need any clean images, in situations where they are available, they cannot utilize the valuable information available in the clean images.

\begin{figure}[t]
    \centering
    \setlength{\tabcolsep}{1pt}
    \begin{tabular}{ccc}
        \textbf{Ground truth} & \textbf{Noisy input} & \textbf{Our output} \\
         \includegraphics[trim={100 0 50 0}, clip, width=2.75cm]{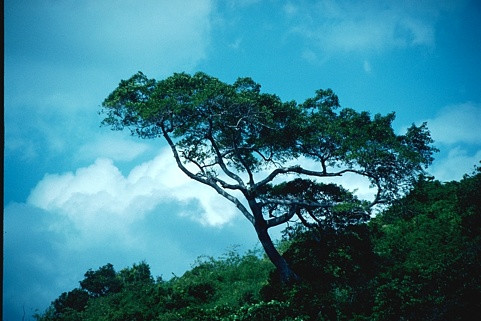} &
         \includegraphics[trim={100 0 50 0}, clip, width=2.75cm]{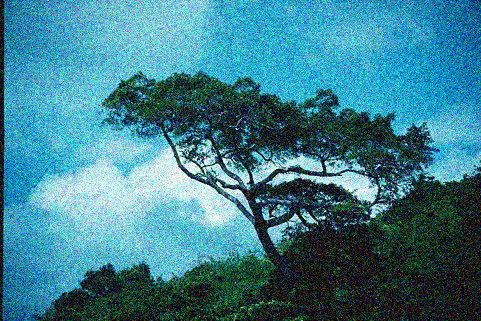} &
         \includegraphics[trim={100 0 50 0}, clip, width=2.75cm]{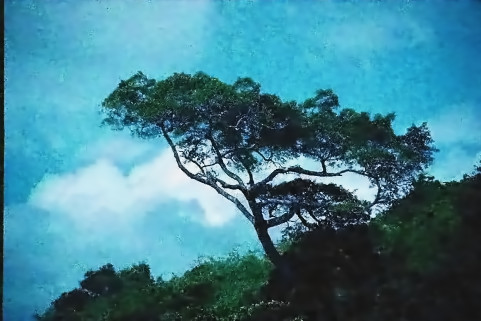}
    \end{tabular}
    \caption{Sample result from our method. Observe that the fine details in the tree are restored without any noticeable blur even when the noise level in the input is high ($\sigma = 35$). Image taken from BSD68 \cite{roth2009fields}}
    \label{fig:intro}
\end{figure}

\noindent Another important class of methods are prior based. Priors are crucial for obtaining a reasonable answer out of all the possible solutions for an ill-posed problem such as image denoising. With handcrafted priors, these methods can be used when clean images are not available. However, these priors have been criticized as they are often chosen for their computational or analytical convenience rather than accuracy. 
\noindent Deep learning has allowed for constructing more accurate priors. Deep image prior \cite{ulyanov2018deep} claims that the architecture of a convolutional neural network alone can act as a prior for natural images. Though the results are good, it is surprising as there is no mathematical justification for why this prior works. Going further, Chen et al \cite{8578431} have used a GAN \cite{goodfellow2014generative} to explicitly construct a prior for realistic noise which they use for denoising. 

\noindent In this paper, we propose an approach for denoising using another class of generative models, called flow-based generative models \cite{DBLP:journals/corr/DinhKB14, DBLP:conf/iclr/DinhSB17}. These models learn an invertible transformation from a complex distribution like images to a simple one like the Gaussian distribution. They have been successfully used to generate realistic images. An example of this is the work by Kingma et al., in \cite{kingma2018glow} where they use flow-based models to generate photorealistic face images using the CelebA HQ dataset \cite{liu2015faceattributes}. Unlike GANs, Flow-based models can explicitly and accurately capture the likelihood function of clean images. As a consequence, they are excellent candidates for learning a realistic prior which is essential for superior denoising performance. Also, they do not suffer from the unstable training dynamics that GANs are notorious for. 

\noindent \autoref{fig:intro} shows a sample result from our method. We train a flow-model on clean images alone while a different network is trained to denoise using only the likelihood specified by the flow-based model. As a result, our method can be used even when there is no pairing between noisy and clean images.

\noindent The main contributions of our work are as follows:
\begin{enumerate}
    \item To the best of our knowledge, this is the first approach to use a flow-based model as a prior for image denoising.
    \item Through extensive experimentation we show that our method has comparable quantitative, as well as, qualitative performance.
\end{enumerate}
\section{Flow-based generative models}
\label{sec:flow}

Flow-based generative models \cite{DBLP:journals/corr/DinhKB14, DBLP:conf/iclr/DinhSB17} learn the bijective transformation from a high-dimensional, complicated random variable $\pmb{X}$ to a latent random variable $\pmb{Z}$. Typically, $\pmb{X}$ represents images in a dataset while $\pmb{Z}$ is assumed to be a standard normal random vector.
\begin{align}
    \pmb{Z} \sim \mathcal{N}(\pmb{0}, \pmb{I}) \\
    \pmb{X} = h(\pmb{Z})
\end{align}
To learn the transformation $h$, the following unbiased estimate of the negative log-likelihood of $\pmb{X}$ is minimized:
\begin{equation}
    \frac{1}{N}\sum_{i=1}^N - \log{P(\pmb{x_i})}
\end{equation}
Here, $\pmb{x_i}$ are samples from the dataset. Using the standard rules of random variable transformation, $\log{P(\pmb{X})}$ can be written as
\begin{equation}
    \log{P(\pmb{X})} = \log{P(\pmb{Z})} - \log{\left|\frac{\mathrm{d}h}{\mathrm{d}\pmb{x}}\right|} \\
\end{equation}
where $\left|\frac{\mathrm{d}h}{\mathrm{d}\pmb{x}}\right|$ is the determinant of the Jacobian of $h$. This term can be further decomposed when $h$ is a composition of several other functions as is typical in a deep neural network.
\begin{align}
    \pmb{X} &= \pmb{Z_0} \xrightarrow{h_1} \pmb{Z_1} \xrightarrow{h_2} \pmb{Z_2} \hdots \xrightarrow{h_n} \pmb{Z_n} = \pmb{Z} \\
    \log{P(\pmb{X})} &= \log{P(\pmb{Z})} - \sum_{i=1}^n\log{\left|\frac{\mathrm{d}\pmb{Z_i}}{\mathrm{d}\pmb{Z_{i - 1}}}\right|} \label{eq:6}
\end{align}
To make the computation of the right hand side of \eqref{eq:6} tractable, flow-based models restrict the class of transformations to those for which the Jacobian is a triangular (or even a diagonal) matrix. A simple example is the following additive coupling layer \cite{DBLP:journals/corr/DinhKB14}:
\begin{align}
    \pmb{y_{p_1}} &= \pmb{x_{p_1}} \\
    \pmb{y_{p_2}} &= \pmb{x_{p_2}} + m(\pmb{x_{p_1}})
\end{align}
where $\pmb{x}, \pmb{y}$ are the inputs and outputs of the layer respectively; $p_1, p_2$ is a partition of the features along the channel dimension and $m$ is an arbitrary transformation. For this layer, it is easy to see that the Jacobian is
\begin{equation}
\begin{bmatrix}
\pmb{I_{p_1}} & 0 \\
\frac{\mathrm{d}m(\pmb{x_{p_1}})}{\mathrm{d}\pmb{x_{p_1}}} &  \pmb{I_{p_2}}
\end{bmatrix}
\label{eq:9}
\end{equation}
where $\pmb{I_{p_1}}, \pmb{I_{p_2}}$ are identity matrices that are of the same size as the partitions $p_1, p_2$. Conveniently, the determinant of the matrix in \eqref{eq:9} is simply 1 and hence it is ideal for use in a flow-based model. Unlike in \cite{DBLP:journals/corr/DinhKB14, DBLP:conf/iclr/DinhSB17, kingma2018glow}, we do not require invertible transformations as there is no need for sampling when we are only learning a prior. Nevertheless, in our work we use the layers and formulation of flow-based models proposed in \cite{kingma2018glow}.

\section{Proposed method}
\label{sec:method}

\begin{figure}[ht]
    \centering
    \includegraphics[width=8cm]{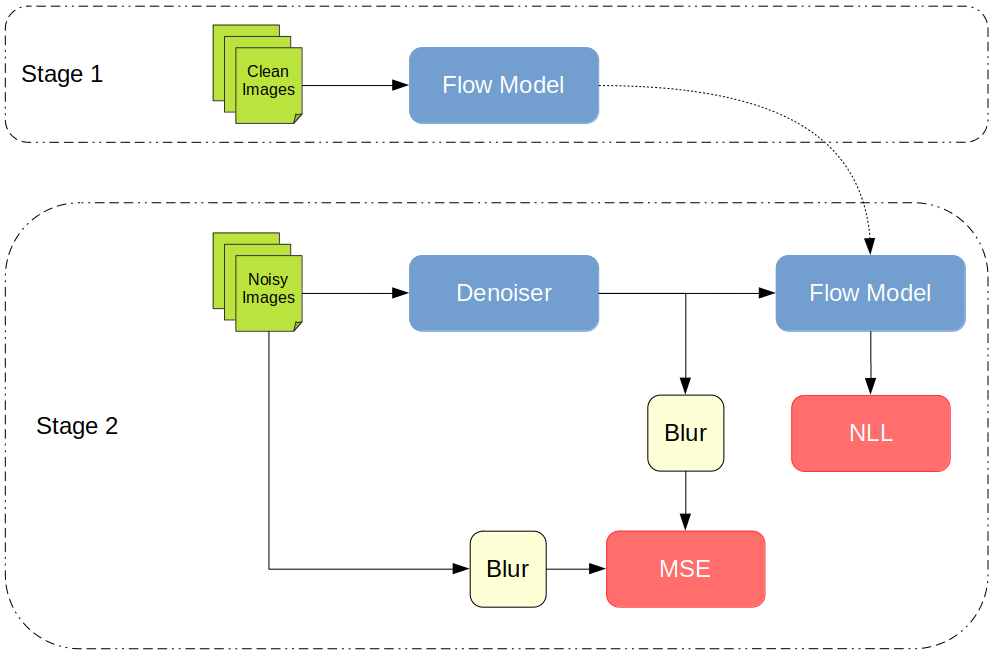}
    \caption{An illustration of our method. In the first stage, we train a flow-based model to learn a prior distribution on clean images. Next, we use this prior along with weak supervision (see subsection 3.2)  to train a denoising network.}
    \label{fig:flow-denoising}
\end{figure}

In this section we describe our two-stage approach (illustrated in \autoref{fig:flow-denoising}) to using the log-likelihood in \eqref{eq:6} as a prior for image denoising.

\subsection{Stage 1: Training the Flow model}

\noindent First, we train a flow-based model based on clean images to learn a transformation from clean images to the standard multivariate Gaussian random variable. Due to structure of the flow-based model as described in \eqref{sec:flow} and the tractable probability density of a Gaussian random variable, we can evaluate \eqref{eq:6} for any given image and obtain the likelihood that the image is clean. 

\noindent Concretely, we train a flow-based model $h$ to minimize the following objective:
\begin{align}
    -\log{P(\pmb{X})} &= -\log{P(\pmb{Z})} + \sum_{i=1}^n\log{\left|\frac{\mathrm{d}\pmb{Z_i}}{\mathrm{d}\pmb{Z_{i - 1}}}\right|}  \\
    &= \frac{1}{2}\|\pmb{Z}\|_2^2 + \sum_{i=1}^n\log{\left|\frac{\mathrm{d}\pmb{Z_i}}{\mathrm{d}\pmb{Z_{i - 1}}}\right|} + C \label{eq:11}
\end{align}
where $C$ is a constant that normalizes the Gaussian distribution. It has no bearing on the training and hence can be eliminated. Note that once the training in Stage 1 is complete, $h$ is fixed during Stage 2.

\subsection{Stage 2: Training the Denoiser}
\noindent Given a noisy image $\pmb{Y}$, the posterior distribution for the corresponding clean image $\pmb{X}$ is
\begin{equation}
    P(\pmb{X}~|~\pmb{Y}) = \frac{P(\pmb{Y}~|~\pmb{X})P(\pmb{X})}{P(\pmb{Y})}
\end{equation}
To obtain the maximum a posteriori (MAP) estimate of the clean image, the denominator can be ignored and the numerator or equivalently its $\log $ value is maximized. 
\begin{equation}
    \argmax_{\pmb{X}} \log{P(\pmb{X}~|~\pmb{Y})} = \argmax_{\pmb{X}} \log{P(\pmb{Y}~|~\pmb{X})} + \log{P(\pmb{X})} \label{eq:13} \\
\end{equation}
\noindent Assuming additive white Gaussian noise, $\log{P(\pmb{Y} | \pmb{X})}$ is simply the negative of the squared error between $\pmb{Y}$ and $\pmb{X}$. Using the flow model $h$ trained in Stage 1, we can also compute the prior log-likelihood of $\pmb{X}$. 

\noindent Based on \eqref{eq:13}, we can formulate a loss function (note the change of signs as by convention, we want to minimize this loss) for the denoiser $d$ as follows:
\begin{equation}
    (\pmb{Y} - \pmb{X})^2 - \lambda \log{P(\pmb{X})} \label{eq:14}
\end{equation}
where $\lambda$ is a hyperparameter that controls the relative importance of the conditional and the prior probability distributions. To be mathematically precise, $\lambda$ depends on the noise level in the image. From our experiments, we also find that the performance of the denoiser is very sensitive to the choice of $\lambda$. This poses a challenge as we want to train a single denoiser for a range of noise levels.

\noindent To reduce the dependency of $\lambda$ on the noise level, we modify the first term in \eqref{eq:14} to instead measure the squared error between blurred versions of $\pmb{X}$ and $\pmb{Y}$. Intuitively speaking, we are training $d$ to copy only the low frequency information from the input $\pmb{Y}$ while adding details that make the output $\pmb{X}$ to look more clean. The flow model $h$ dictates what details are added to $\pmb{Y}$.

\noindent The final form of the loss function we use for the denoiser $d$ is
\begin{equation}
(B(\pmb{Y}) - B(\pmb{X}))^2 - \lambda \log{P(\pmb{X})} \label{eq:15}
\end{equation}
Here $B$ is a local mean filter, the size of which is chosen to be $3\times 3$, as that gave the best performance on the validation set.
\section{Experiments}
\label{sec:experiments}

\subsection{Training Details}
\noindent We use the validation set of MS COCO \cite{coco} for our training. Of the 41K images it contains, we use a subset of 20K images as our clean image dataset. We add Gaussian noise to another subset of 20K images to form our noisy image dataset.  As we want our method to be agnostic to noise level, for each image, the standard deviation of the added noise is chosen uniformly in the interval $[0, 50]$. We set aside the remaining 1K images for validation to tune the hyperparameters $\lambda$ and the size of the local mean filter.

\subsubsection{Stage 1}
We use the architecture described in \cite{kingma2018glow} for the flow-based model. We feed patches of size 32 from images in the clean dataset as input to this model. Using the loss in \eqref{eq:11}, we train for 100 epochs using the Adam optimizer \cite{kingma2014adam} with learning rate $ = 1 \times 10^{-3}$, $\beta_1 = 0.9$, $\beta_2 = 0.999$.

\subsubsection{Stage 2}
We use the ResNet \cite{He_2016_CVPR} for our denoiser. Because the flow-based model only accepts fixed size inputs and the ResNet does not change input size, we use input patches of size 32. In this stage, however, they are extracted from noisy images. Using \eqref{eq:15}, we train only the denoiser, for 100 epochs using the Adam optimizer with the same parameter settings as in stage 1. We experimented with various choices of $\lambda$ and the size of the local mean filter. Based on our results, we choose $\lambda = 1.5 \times 10^{-6}$ and local mean filter of size $3 \times 3$ as they give the best PSNR values.

\subsection{Results}

\begin{figure*}[t]
    \centering
    \begin{tabular}{cccccc}
    Ground truth & Noisy input & BM3D & Noise2Void & Deep image prior & Ours \\
    \includegraphics[width=2.5cm]{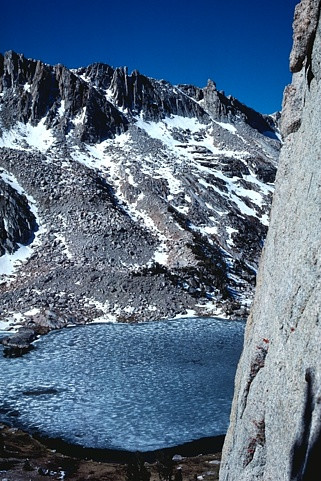} &
    \includegraphics[width=2.5cm]{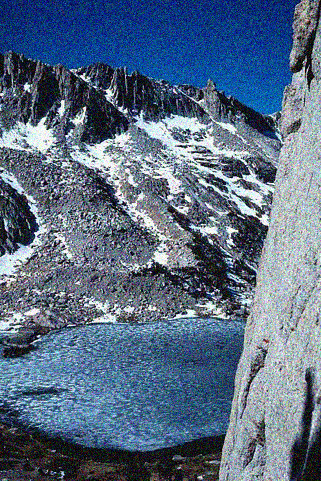} &
    \includegraphics[width=2.5cm]{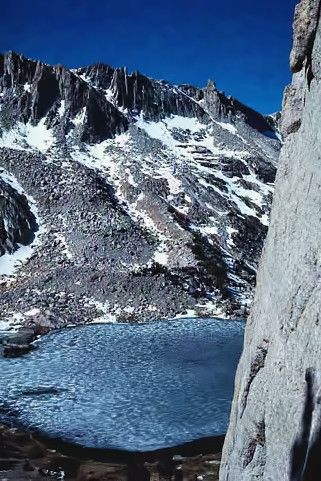} &
    \includegraphics[width=2.5cm]{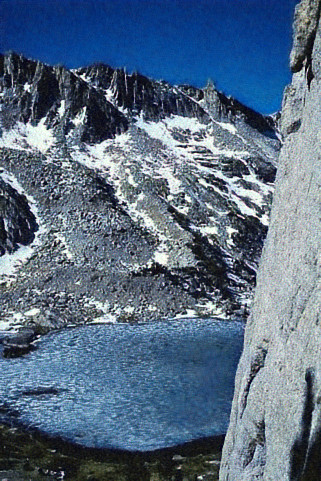} &
    \includegraphics[width=2.5cm]{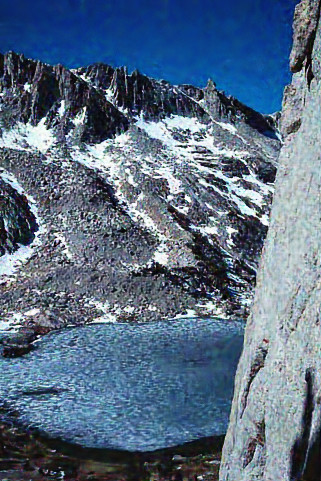} &
    \includegraphics[width=2.5cm]{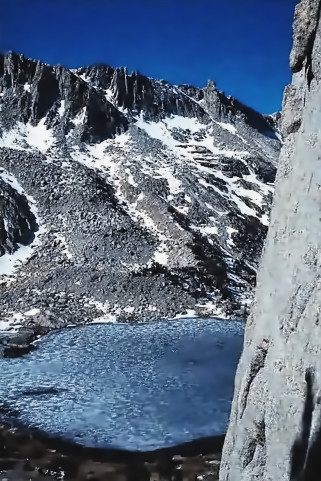} \\
    
    \includegraphics[width=2.5cm]{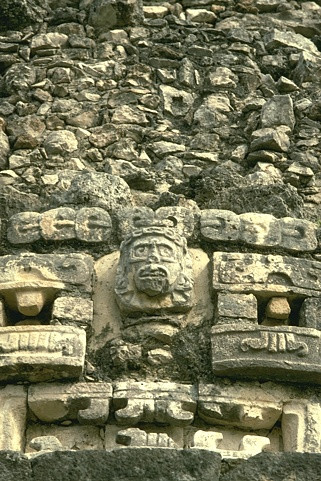} &
    \includegraphics[width=2.5cm]{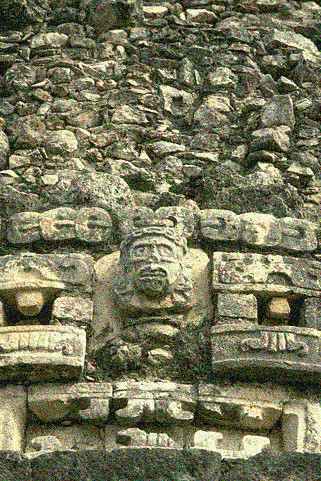} &
    \includegraphics[width=2.5cm]{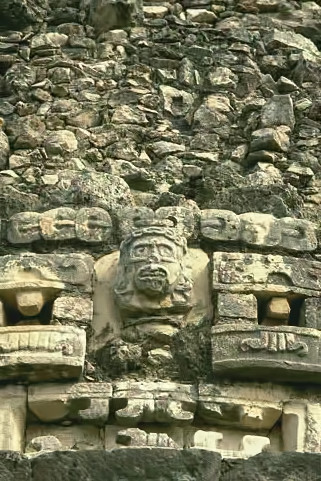} &
    \includegraphics[width=2.5cm]{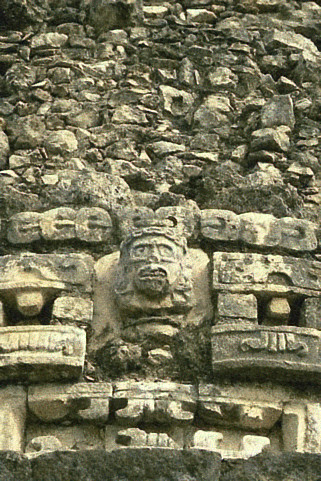} &    \includegraphics[width=2.5cm]{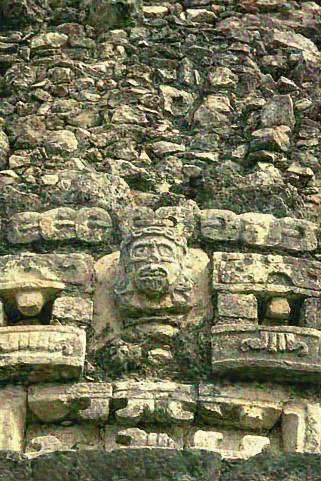} &
    \includegraphics[width=2.5cm]{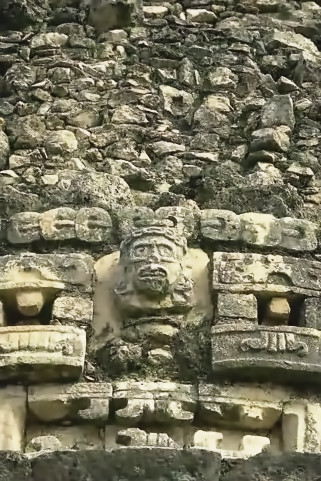} \\
    
    \includegraphics[width=2.5cm]{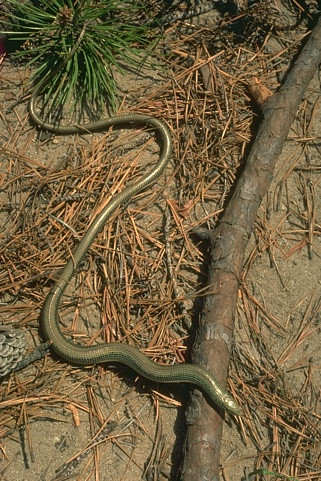} &
    \includegraphics[width=2.5cm]{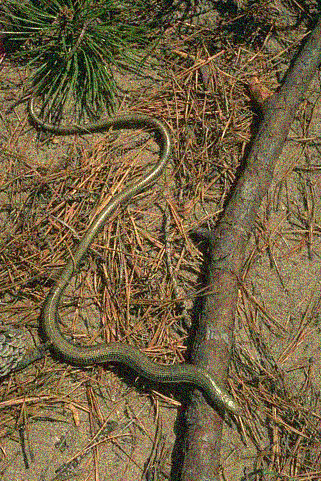} &
    \includegraphics[width=2.5cm]{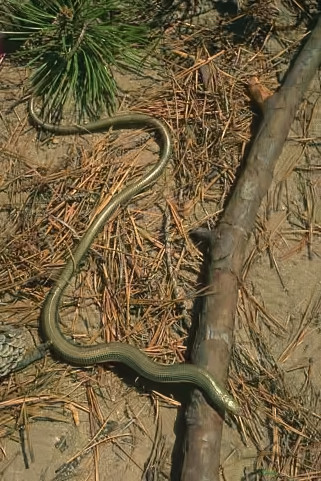} &
    \includegraphics[width=2.5cm]{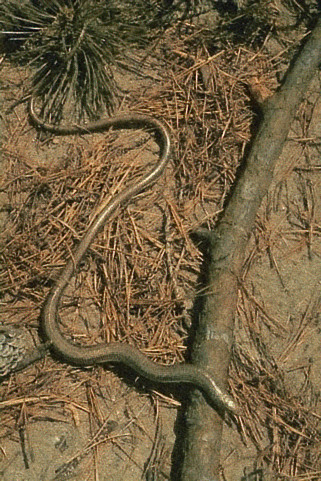} &
    \includegraphics[width=2.5cm]{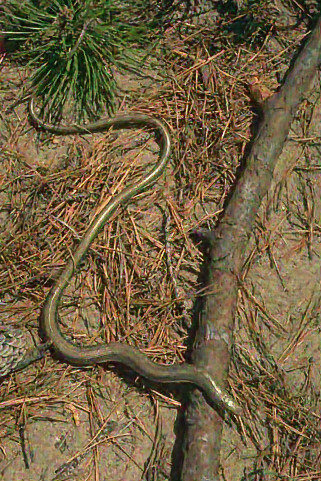} &
    \includegraphics[width=2.5cm]{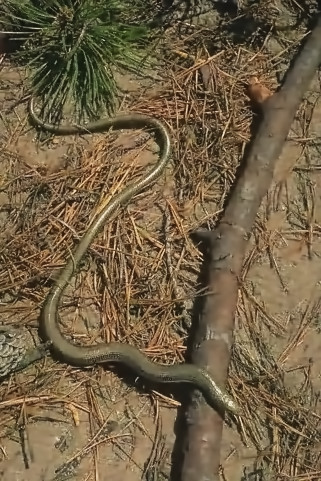} \\
    \end{tabular}
    \caption{Qualitative results. Here, we show the ground truth, the noisy input (Gaussian noise, $\sigma=25$) and the denoised outputs from BM3D \cite{4271520}, Noise2Void \cite{8954066}, Deep image prior \cite{ulyanov2018deep} and finally, our method. None of these methods need supervision. }
    \label{fig:quant-results}
\end{figure*}

\noindent Following \cite{4271520, 8954066}, we evaluate our method on the BSD68 dataset \cite{roth2009fields} for different noise levels and compare it with BM3D \cite{4271520}, Noise2Void \cite{8954066}, Deep image prior \cite{ulyanov2018deep}. All comparisons are made using either results reported in the respective papers or those obtained from running the code that the authors have generously shared. \autoref{tab:1} shows the average PSNR values of different methods for images from BSD68. Although PSNR is not an accurate metric for perceptual quality, our method performs competitively with Noise2Void and is better than Deep image prior.

\begin{table}[htbp]
\centering
\begin{tabular}{|c|c|c|c|c|c|}
    \hline
     Method & BM3D & DnCNN & N2V & DIP & Ours\\
     \hline\hline
     $\sigma=$ 15 & 33.14 & 31.73 & 28.92 & 27.58 & 29.10 \\
     \hline
     $\sigma=$ 25 & 30.22 & 29.23 & 27.68 & 26.6 & 28.61 \\
     \hline
     $\sigma=$ 35 & 28.25 & 28.95 & 26.51 & 25.97 & 26.2 \\
     \hline
\end{tabular}
\caption{Quantitative results. We show PSNR (dB) of various denoising methods, namely, BM3D \cite{4271520}, DnCNN \cite{7839189}, Noise2Void \cite{8954066}, Deep image prior \cite{ulyanov2018deep} and our method. Of these, only DnCNN is fully supervised.}
\label{tab:1}
\end{table}

\noindent \autoref{fig:quant-results} shows qualitative comparison of our results with other methods. Our method is able to remove noise effectively without blurring any textures, details or sharp edges (this is obvious in the sky in the first set of images). Deep image prior produces outputs that still have visible noise. Noise2Void, although better than Deep image prior, fails in some cases. An example of this is the blades of grass in the third set of images where the output from Noise2Void is noticeably desaturated.


\section{Conclusions}
\label{sec:conclusions}
\noindent We have proposed the use of flow-based model as a mathematically justifiable and realistic prior for image denoising. We have conducted qualitative and quantitative experiments on the BSD68 \cite{roth2009fields} dataset that reveals the competitive performance of our method.

\noindent Motivated by our success, we conjecture that using a flow-based model prior should be effective for solving other image restoration tasks such as image deblurring and super-resolution in an unsupervised fashion.

\bibliographystyle{IEEEbib}
\bibliography{kattakinda}

\begin{thebibliography}{10}

\bibitem{4271520}
K.~{Dabov}, A.~{Foi}, V.~{Katkovnik}, and K.~{Egiazarian},
\newblock ``Image denoising by sparse 3-d transform-domain collaborative
  filtering,''
\newblock {\em IEEE Transactions on Image Processing}, vol. 16, no. 8, pp.
  2080--2095, Aug 2007.

\bibitem{dong2012nonlocally}
Weisheng Dong, Lei Zhang, Guangming Shi, and Xin Li,
\newblock ``Nonlocally centralized sparse representation for image
  restoration,''
\newblock {\em IEEE transactions on Image Processing}, vol. 22, no. 4, pp.
  1620--1630, 2012.

\bibitem{gu2014weighted}
Shuhang Gu, Lei Zhang, Wangmeng Zuo, and Xiangchu Feng,
\newblock ``Weighted nuclear norm minimization with application to image
  denoising,''
\newblock in {\em Proceedings of the IEEE conference on computer vision and
  pattern recognition}, 2014, pp. 2862--2869.

\bibitem{zhang2017learning}
Kai Zhang, Wangmeng Zuo, Shuhang Gu, and Lei Zhang,
\newblock ``Learning deep cnn denoiser prior for image restoration,''
\newblock in {\em Proceedings of the IEEE conference on computer vision and
  pattern recognition}, 2017, pp. 3929--3938.

\bibitem{7839189}
K.~{Zhang}, W.~{Zuo}, Y.~{Chen}, D.~{Meng}, and L.~{Zhang},
\newblock ``Beyond a gaussian denoiser: Residual learning of deep cnn for image
  denoising,''
\newblock {\em IEEE Transactions on Image Processing}, vol. 26, no. 7, pp.
  3142--3155, July 2017.

\bibitem{lefkimmiatis2017non}
Stamatios Lefkimmiatis,
\newblock ``Non-local color image denoising with convolutional neural
  networks,''
\newblock in {\em Proceedings of the IEEE Conference on Computer Vision and
  Pattern Recognition}, 2017, pp. 3587--3596.

\bibitem{pmlr-v80-lehtinen18a}
Jaakko Lehtinen, Jacob Munkberg, Jon Hasselgren, Samuli Laine, Tero Karras,
  Miika Aittala, and Timo Aila,
\newblock ``{N}oise2{N}oise: Learning image restoration without clean data,''
\newblock in {\em Proceedings of the 35th International Conference on Machine
  Learning}, Jennifer Dy and Andreas Krause, Eds., Stockholmsmässan, Stockholm
  Sweden, 10--15 Jul 2018, vol.~80 of {\em Proceedings of Machine Learning
  Research}, pp. 2965--2974, PMLR.

\bibitem{8954066}
A.~{Krull}, T.~{Buchholz}, and F.~{Jug},
\newblock ``Noise2void - learning denoising from single noisy images,''
\newblock in {\em 2019 IEEE/CVF Conference on Computer Vision and Pattern
  Recognition (CVPR)}, June 2019, pp. 2124--2132.

\bibitem{roth2009fields}
Stefan Roth and Michael~J Black,
\newblock ``Fields of experts,''
\newblock {\em International Journal of Computer Vision}, vol. 82, no. 2, pp.
  205, 2009.

\bibitem{ulyanov2018deep}
Dmitry Ulyanov, Andrea Vedaldi, and Victor Lempitsky,
\newblock ``Deep image prior,''
\newblock in {\em Proceedings of the IEEE Conference on Computer Vision and
  Pattern Recognition}, 2018, pp. 9446--9454.

\bibitem{8578431}
J.~{Chen}, J.~{Chen}, H.~{Chao}, and M.~{Yang},
\newblock ``Image blind denoising with generative adversarial network based
  noise modeling,''
\newblock in {\em 2018 IEEE/CVF Conference on Computer Vision and Pattern
  Recognition}, June 2018, pp. 3155--3164.

\bibitem{goodfellow2014generative}
Ian Goodfellow, Jean Pouget-Abadie, Mehdi Mirza, Bing Xu, David Warde-Farley,
  Sherjil Ozair, Aaron Courville, and Yoshua Bengio,
\newblock ``Generative adversarial nets,''
\newblock in {\em Advances in neural information processing systems}, 2014, pp.
  2672--2680.

\bibitem{DBLP:journals/corr/DinhKB14}
Laurent Dinh, David Krueger, and Yoshua Bengio,
\newblock ``{NICE:} non-linear independent components estimation,''
\newblock in {\em 3rd International Conference on Learning Representations,
  {ICLR} 2015, San Diego, CA, USA, May 7-9, 2015, Workshop Track Proceedings},
  Yoshua Bengio and Yann LeCun, Eds., 2015.

\bibitem{DBLP:conf/iclr/DinhSB17}
Laurent Dinh, Jascha Sohl{-}Dickstein, and Samy Bengio,
\newblock ``Density estimation using real {NVP},''
\newblock in {\em 5th International Conference on Learning Representations,
  {ICLR} 2017, Toulon, France, April 24-26, 2017, Conference Track
  Proceedings}. 2017, OpenReview.net.

\bibitem{kingma2018glow}
Durk~P Kingma and Prafulla Dhariwal,
\newblock ``Glow: Generative flow with invertible 1x1 convolutions,''
\newblock in {\em Advances in Neural Information Processing Systems}, 2018, pp.
  10215--10224.

\bibitem{liu2015faceattributes}
Ziwei Liu, Ping Luo, Xiaogang Wang, and Xiaoou Tang,
\newblock ``Deep learning face attributes in the wild,''
\newblock in {\em Proceedings of International Conference on Computer Vision
  (ICCV)}, December 2015.

\bibitem{coco}
Tsung-Yi {Lin}, Michael {Maire}, Serge {Belongie}, Lubomir {Bourdev}, Ross
  {Girshick}, James {Hays}, Pietro {Perona}, Deva {Ramanan}, C.~Lawrence
  {Zitnick}, and Piotr {Doll{\'a}r},
\newblock ``{Microsoft COCO: Common Objects in Context},''
\newblock {\em arXiv e-prints}, p. arXiv:1405.0312, May 2014.

\bibitem{kingma2014adam}
Diederik~P Kingma and Jimmy Ba,
\newblock ``Adam: a method for stochastic optimization. corr abs/1412.6980
  (2014),'' 2014.

\bibitem{He_2016_CVPR}
Kaiming He, Xiangyu Zhang, Shaoqing Ren, and Jian Sun,
\newblock ``Deep residual learning for image recognition,''
\newblock in {\em The IEEE Conference on Computer Vision and Pattern
  Recognition (CVPR)}, June 2016.

\end{thebibliography}

\end{document}